\documentclass{svmult}

\usepackage{makeidx}         
\usepackage{graphicx}        
\usepackage{multicol}        
\usepackage{amssymb}
\usepackage{amsmath}
\usepackage[bottom]{footmisc}

\makeindex             

\begin{document}

\title*{Dynamic Process of Money Transfer Models}
\author{Yougui Wang\inst{*}\and
Ning Ding}
\institute{Department of Systems Science, School of Management,
Beijing Normal University, Beijing, 100875, People's Republic of
China \texttt{ygwang@bnu.edu.cn} }
%
%
\maketitle
\begin{abstract}
We have studied numerically the statistical mechanics of the
dynamic phenomena, including money circulation and economic
mobility, in some transfer models. The models on which our
investigations were performed are the basic model proposed by A.
Dr\u{a}gulescu and V. Yakovenko \cite{model1}, the model with
uniform saving rate developed by A. Chakraborti and B.K.
Chakrabarti \cite{model2}, and its extended model with diverse
saving rate \cite{model3}. The velocity of circulation is found to
be inversely related with the average holding time of money. In
order to check the nature of money transferring process in these
models, we demonstrated the probability distributions of holding
time. In the model with uniform saving rate, the distribution
obeys exponential law, which indicates money transfer here is a
kind of Poisson process. But when the saving rate is set
diversely, the holding time distribution follows a power law. The
velocity can also be deduced from a typical individual's optimal
choice. In this way, an approach for building the micro-foundation
of velocity is provided. In order to expose the dynamic mechanism
behind the distribution in microscope, we examined the mobility by
collecting the time series of agents' rank and measured it by
employing an index raised by economists. In the model with uniform
saving rate, the higher saving rate, the slower agents moves in
the economy. Meanwhile, all of the agents have the same chance to
be the rich. However, it is not the case in the model with diverse
saving rate, where the assumed economy falls into stratification.
The volatility distribution of the agents' ranks are also
demonstrated to distinguish the differences among these models.

\end{abstract}

\section{Introduction}
\label{intro}

Recently, wealth or income distribution has attracted much
attention in the field of econophysics \cite{solomon,bouch,sca}.
More than 100 years ago, Italian economist Pareto first found that
the income distribution follows an universal power law
\cite{pareto}. However, the economy has undergone dramatic
transitions in last century, some researchers had doubted about if
the law still holds in the modern stage and turned to reexamine
the income distribution and its shift by employing income tax data
\cite{emp1,emp2,emp3,emp4,emp5}. The empirical analysis showed
that in many countries the income distribution typically presents
with a power-law tail, and majority of the income distribution can
be described by an exponential law. This universal shape of
distribution and its shift trigger an increasing interests in
exploring the mechanism behind them. To solve this problem,
several multi-agent models have been developed by applying
principles of statistical mechanics \cite{model1,model2,model3,
model4,follow,slanina}. In these models, economic system is
analogized to the ideal gas, where the agents can be regarded as
particles, and money is just like energy. Therefore, the trading
between agents can be viewed as collisions between particles in
the gas. By using such analogy, the developed approach that
applied to the ideal gas system now can be used to study this kind
of economic system. Whatever the trading rule is set in these
models, it is worthy noting that money is always transferred from
one agent to another in the trading process. So this kind of
models could be referred as money transfer models \cite{transfer}.

Leading the search into this issue was a paper by A.
Dr\u{a}gulescu and V. Yakovenko \cite{model1}. In their ideal-gas
model, the economy is closed and the amount of money transferred
in each round of trading is determined randomly. Their simulation
analysis shows that the steady money distribution follows an
exponential law. Several papers have extended the work by
introducing different characteristics into the model and found
that different trading rule may lead to different shapes of money
distribution. A. Chakraborti and B.K. Chakrabarti examined the
case where the agents do not take out all amount of money as they
participate the exchange, but instead they save a part of their
money \cite{model2}. This case is well grounded in reality, and
the ratio they save is called saving rate hereafter. When the
saving rate are the same for all agents, the money distribution
obeys a Gamma law \cite{analysis1}. However, when the agents'
saving rates are set randomly, the money distribution changes to a
Power-law type \cite{model3}. A second extension looks at
non-conservation. F. Slanina considered a case that the economy is
not conserved but opened, and so he regarded it as inelastic
granular gases \cite{slanina}. Some further studies manage to seek
for the exact mathematical solution by using a master equation
\cite{analysis2,analysis3}.

In fact, money transfer is a dynamic process. Besides the money
distribution, it possess some other presentations. Thus,
investigating the distribution only can not provide the whole
picture of the relationship between the distribution and the
trading rule. Some efforts have been put into the study on the
dynamic mechanism behind the distribution, that opens more windows
to observe how the economy works.

These works can be divided into two parts. One is about how the
money moves in the assumed economy
\cite{circulation1,circulation2,velocity}. As we know, the money
is not static even after the money distribution gets steady. They
are always transferred among agents. Naturally, because of the
randomness, whether in the simulations or in the reality, the time
interval that money stays in one agent's pocket is a random
variable which is named as holding time. The introduction of
holding time opens a new path to understanding of the circulation
velocity at micro level.

The other one is about how agents' positions shift in the economy
\cite{mob}. Like the money, agents are not static in the
transferring process. If the agents are sorted according to the
amount of money they hold, it is found that the rank of any agent
varies over time. This phenomenon is called mobility in economics.
According to economists' argument, only analysis on the
distribution is not sufficient especially when comparing the
generating mechanism of income and the
inequality\cite{kuznets,jj}.

In addition, the study on the dynamic characters in the proposed
models makes the evaluation criteria more complete. The aim of
econophysicists to develop these models is to mimic the real
economy by abstracting its essence. However, we cannot judge
whether such abstraction is reasonable or not depending on the
shape of distribution only. Thus, we must take the circulation and
mobility into account when constructing a ``good" multi-agent
model.

In this paper, the dynamic processes of the transfer models are
investigated by examining the holding time distribution and the
degree of mobility. The models and simulations will be briefly
presented in the next section. In the Sec. 3 and 4, we will show
the nature of circulation and mobility in these models
respectively. Finally, we will give our conclusion in Sec. 5.

\section{Models and Simulations}

We start with the transfer model proposed by A. Dr\u{a}gulescu and
V. Yakovenko, in which the economic system is closed, put it in
another way, the total amount of money $M$ and the number of
economic agents $N$ are fixed. Each of agents has a certain amount
of money initially. In each round of trading process, two agents
$i,j$ are chosen to take part in the trade randomly. And it is
also decided randomly which one is the payer or receiver. Suppose
the amounts of money held by agent $i$ and $j$ are $m_i$ and
$m_j$, the amount of money to be exchanged $\Delta m$ is decided
by the following trading rule:
\begin{equation}\label{basic}
    \Delta m=\frac{1}{2}\varepsilon (m_i+m_j),
\end{equation}
where $\varepsilon$ is a random number from zero to unit. If the
payer cannot afford the money to be exchanged, the trade will be
cancelled. This model is very simple and extensible which is named
as the basic model in this paper.

When A. Chakraborti and B.K. Chakrabarti intended to extend the
basic model, they argued that the agents always keep some of money
in hand as saving when trading. The ratio of saving to all of the
money held is denoted by $s$ and called saving rate in this paper.
For all the agents, the saving rates are set equally before the
simulations. Like the trading pattern of the basic model, two
agents $i,j$ are chosen out to participate the trade in each
round. Suppose that at $t$-th round, agents $i$ and $j$ take part
in trading, so at $t+1$-th round their money $m_i(t)$ and $m_j(t)$
change to
\begin{equation}\label{aaa}
m_i(t+1)=m_i(t)+\Delta m; m_j(t+1)=m_j(t)-\Delta m,
\end{equation}
where \begin{equation}\label{deltam1}
    \Delta m=(1-s)[(\varepsilon-1)m_i(t)+\varepsilon
    m_j(t)],
\end{equation}
and $\varepsilon$ is a random fraction. It can be seen that
$\Delta m$ might be negative. That means agent $i$ is probably the
payer of the trade. This model degenerates into the basic model if
$s$ is set to be zero. In this model, all of agents are homogenous
with the same saving rate. So we call it the model with uniform
saving rate.

This model was further developed by B.K. Chakrabarti's research
group by setting agents' saving rates randomly before the
simulations and keeping them unchanged all through the
simulations. Likewise, this is called the model with diverse
saving rate. Correspondingly, the trading rule Equation
(\ref{deltam1}) changes to
\begin{equation}\label{deltam2}
    \Delta m=(1-s_i)(\varepsilon-1)m_i(t)+(1-s_j)\varepsilon
    m_j(t),
\end{equation}
where $s_i$, $s_j$ are the saving rates of agent $i$ and $j$
respectively.

Our following investigations on the dynamic phenomena is based on
these three models. The scale is the same for all the simulations:
the number of agent $N$ is $1,000$ and the total amount of money
$M$ is $100,000$.

\section{Money Circulation}

As the medium of exchange, money is held and transferred by
people. In the process of money transferring, if an agent receives
money from others at one moment, he will hold it for a period, and
eventually pays it to another agent. The time interval between the
receipt of the money and its disbursement is named as holding
time. We introduce the probability distribution function of
holding time $P_h(\tau)$, which is defined such that the amount of
money whose holding time lies between $\tau$ and $\tau+d\tau$ is
equal to $MP_h(\tau)d\tau$. In the stationary state, the fraction
of money $MP_h(\tau)\,d\tau$ participates in the exchange after a
period of $\tau$. Then the average holding time can be expressed
as
\begin{equation}\label{velocity}
\bar{\tau}=\int\nolimits^\infty_0 P_h(\tau)\,\tau\,d\tau.
\end{equation}
The velocity indicates the speed at which money circulates. Since
money is always spent randomly in exchange, the transferring
process can be deemed as a Poisson type, and the velocity of money
can then be written as \cite{circulation1}
\begin{equation}
V=\frac{1}{\bar{\tau}}.  \label{relation}
\end{equation}
This is the statistical expression of the circulation velocity of
money in terms of holding time.

Two caveats to this conclusion are in order. First, we need to
observe the probability density function of holding time to check
whether the transfer of money is a Poisson process. If the
assumption is correct, the probability density function must take
the following form
\begin{equation}
P(\tau )=\lambda e^{-\lambda \tau },  \label{Gama}
\end{equation}%
where $\lambda $ corresponds to the intensity of the Poisson
process. We have carried out the measurement of holding time in
our previous work \cite{circulation2}. In those simulations, the
time interval between the moments when the money takes part in
trade after $t_0$ for the first two times is recorded as holding
time, supposing we start to record at round $t_0$. The data were
collected after majority of money($>99.9\%$) had been recorded and
over 100 times with different random seeds.

The simulation results are shown in Fig.1. It can be seen the
probability distributions of holding time decay exponentially in
the model with uniform saving rate. This fact indicates that the
process is a Poisson process. On the other case, when the saving
rates are set diversely, the distribution changes to a power-law
type.
\begin{figure}
\centering
\includegraphics[height=5.5cm]{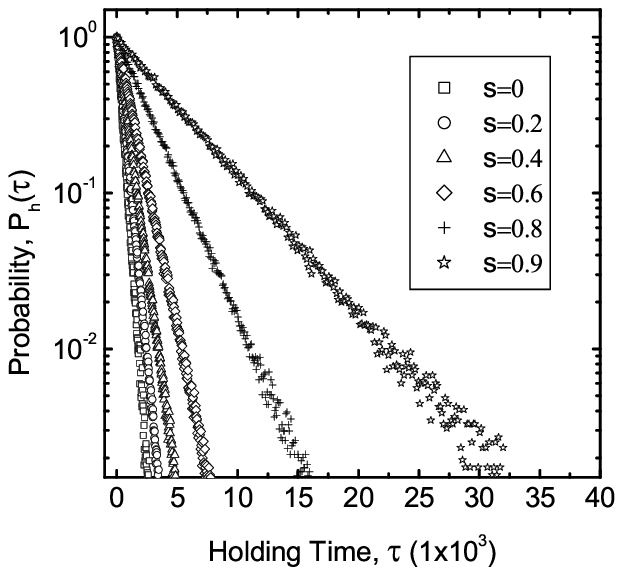}
\includegraphics[height=5.5cm]{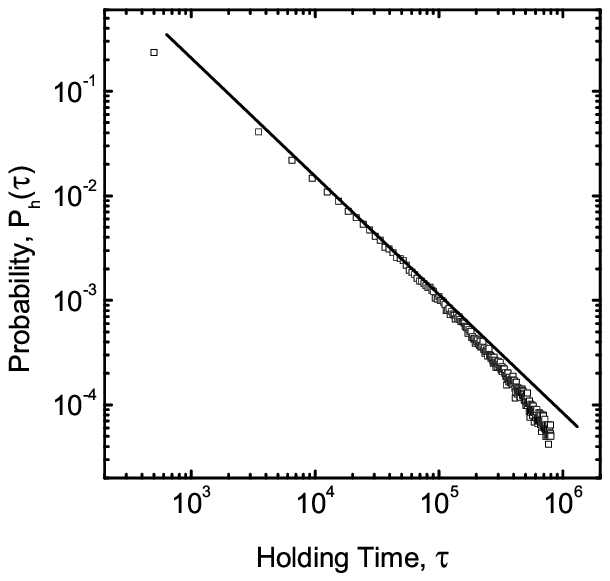}
\caption{The stationary distributions of holding time: (left
panel) for the model with uniform saving rate in a
semi-logarithmic scale, (right panel) for the model with diverse
saving rate in a double-logarithmic scale, where the fitting
exponent of the solid line is about $-1.14$.  Note that in the
figure the probabilities have been scaled by the maximum
probability respectively. }
\label{fig:1}       
\end{figure}

In the model with uniform saving rate, the monetary circulation
velocity corresponds to the intensity of Poisson process, which is
negatively related to the saving rate. Form Fig. 1 we can see that
the lower the saving rate is, the steeper the distribution curve.
This result is also plotted in Fig. 2, which tells us the relation
between the velocity and the saving rate is not linear.

\begin{figure}
\centering
\includegraphics[height=5.5cm]{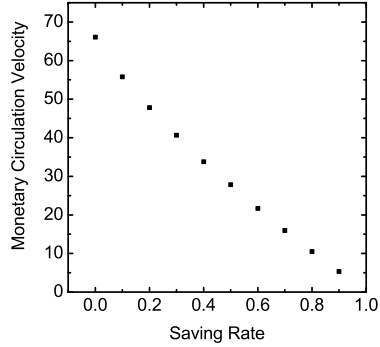}
\caption{The monetary circulation velocity versus the saving rate
in the model with uniform saving rate.}
\label{fig:2}       
\end{figure}

Second, the relation between the velocity of money and the average
holding time suggests that the velocity could be investigated by
examining how economic agents make decisions on the holding time
of money. There are many kinds of agents who may have different
characters when they utilize money in an economic system, such as
consumers, firms, and investors etc. We can choose one of them as
a representative to examine how their spending decisions affect
the velocity. The typical one is consumers whose behavior has
always been depicted by the life-cycle model prevailed in
economics. The model considers a representative individual who
expects to live $T$ years more. His object is to maximize the
lifetime utility
\begin{equation}
U=\int\nolimits_{0}^{T}u(C(t))\,dt,  \label{utility}
\end{equation}
subject to the budget constraint condition
\begin{equation}
\int\nolimits_{0}^{T}C(t)\,dt\leq
W_{0}+\int\nolimits_{0}^{T}Y(t)\,dt, \label{constraint}
\end{equation}
where\ $u(\cdot )$ is an instantaneous concave utility function,
and $C(t)$ is his consumption in time $t$. The individual has
initial wealth of $W_{0}$ and expects to earn labor income $Y(t)$
in the working period of his or her life. The main conclusion
deduced from this optimal problem is that the individual wants to
smooth his consumption even though his income may fluctuate in his
life time. From this conclusion, we can also calculate the average
holding time of money based on the time path of income and
consumption as the following form
\begin{equation}
\bar{\tau}=\frac{\int\nolimits_{0}^{T}[C(t)-Y(t)]t\,dt}{\int%
\nolimits_{0}^{T}Y(t)\,dt}.  \label{delaytime}
\end{equation}
With a few manipulations in a simple version of the life-cycle
model \cite{velocity}, we get
\begin{equation}
V=\frac{2}{T-T_{0}}.  \label{valueofv}
\end{equation}
This result tells us that the velocity of money depends on the
difference between the expected length of life $T$ and that of
working periods $T_{0}$. It also implies that the velocity, as an
aggregate variable, can be deduced from the individual's optimal
choice. In this way, a solid micro foundation for velocity of
money has been constructed.

\section{Economic Mobility}

It is the economists' consensus that static snapshots of income
distribution alone is not sufficient for meaningful evaluation of
wellbeing and the equality. This can be understood easily from a
simple example. Suppose in an economy there are two individuals
with money \$1 and \$2 initially. At the next moment, the amount
of money held by them changes to \$2 and \$1. The distribution in
this case is unchanged, but the ranks of both agents vary over
time. Although the system seems unequal at either of the two
moments in terms of the distribution, the fact is that the two
individuals are quite equal combining these two moments. Besides,
from this simple example, it can also been found that the
structure of economy may vary heavily with an unchanged
distribution. Thus the investigation on mobility is helpful not
only to the measurement on equality but also to the exposure of
the mechanism behind the distribution.

We investigated the mobility in the referred transfer models by
placing emphasis on the ``reranking" phenomenon. To show this kind
of mobility, we sorted all of agents according to their money and
recorded their ranks at the end of each round. All of data were
collected after the money distributions get stationary and the
sampling time interval was set to be 1000 rounds.

\begin{figure}
\includegraphics[height=5.5cm]{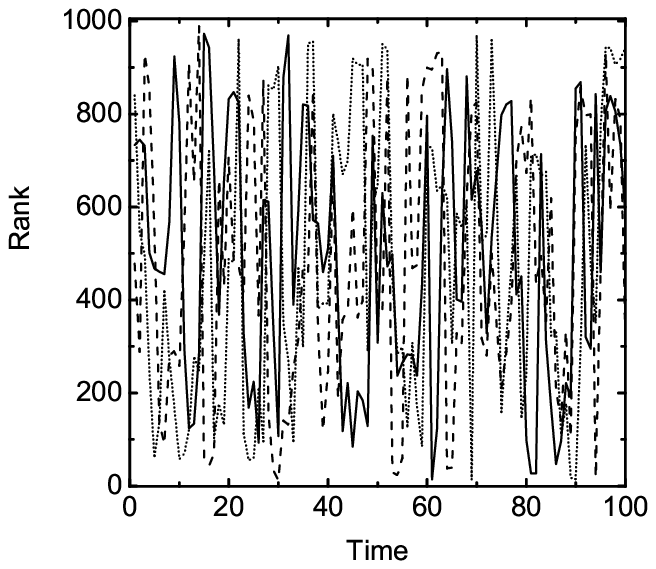}
\includegraphics[height=5.5cm]{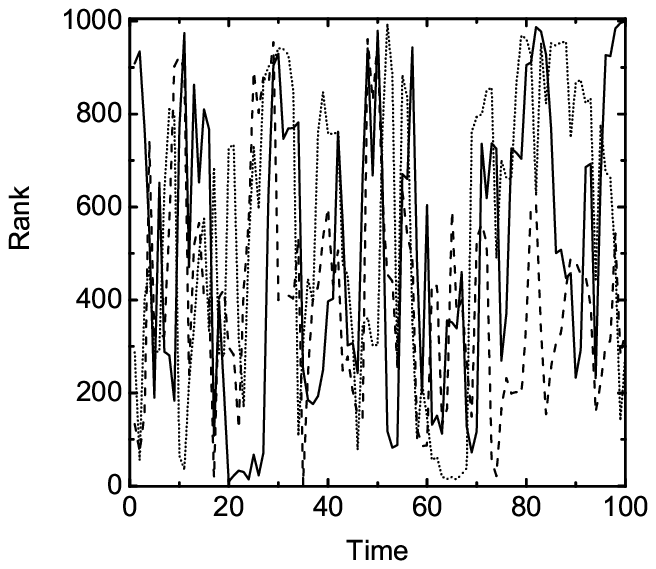}
\begin{center}
\includegraphics[height=5.5cm]{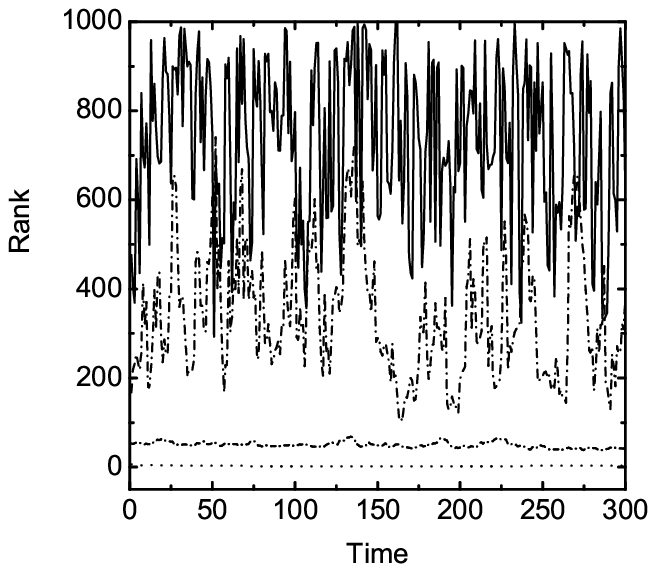}
\end{center}
\caption{The typical time series of rank (a) from basic model, (b)
from the model with uniform saving rate $s=0.5$ and (c) from the
model with diverse saving rate where the saving rates of these
typical agents are 0.99853, 0.9454, 0.71548 and 0.15798 (from
bottom to top ) respectively.}
\label{fig:3}       
\end{figure}

The time series of rank in these three models are shown in Fig.3.
Then, we can compare the characters of rank fluctuation of these
models. All of the agents in the basic model and the model with
uniform saving rate can be the rich and be the poor. The rich have
the probability to be poor and the poor also may be luck to get
money to be the rich. The mobility in these two model are quite
similar except the fluctuation frequency of the time series. The
economy in the model with diverse saving rate is highly stratified
(see Fig. 3c). The rich always keep their position, and the poor
are doomed to be the poor. The agents in each level differ in
their rank fluctuations. The higher the agent' rank, the smaller
the variance of his rank.

\begin{table}
 \caption{Comparison of the Three
Transfer Models in Mobility}\label{a} \centering
\begin{tabular}{c|c|c}
  \hline
    & Mobility $l(t,t')$ & Stratification \\
  \hline
  The Basic Model   & 0.72342 & No \\
  \hline
  The Model with Uniform Saving Rate &   & No \\
  $s=0.1$   &  0.70269 & \\
  $s=0.3$   &  0.65165 & \\
  $s=0.5$   &  0.58129 & \\
  $s=0.7$   &  0.4773 & \\
  $s=0.9$   &  0.30212 & \\
  \hline
  The Model with Diverse Saving Rate & 0.19671 & Yes \\
  \hline
\end{tabular}
\end{table}

\begin{figure}[b]
\includegraphics[height=5.5cm]{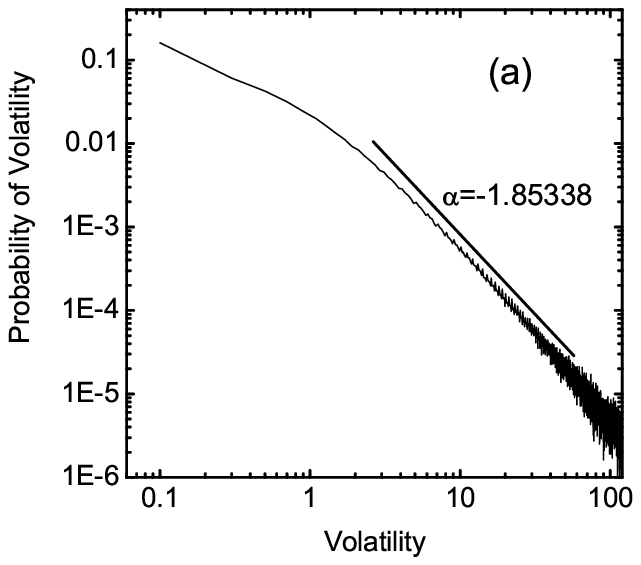}
\includegraphics[height=5.5cm]{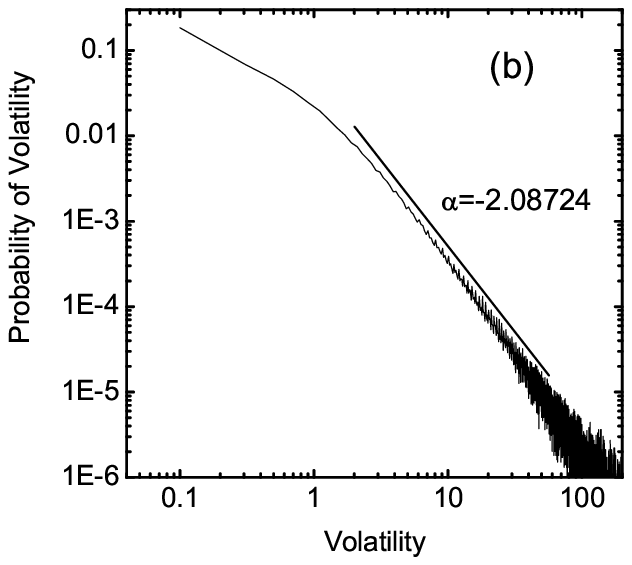}
\begin{center}
\includegraphics[height=5.5cm]{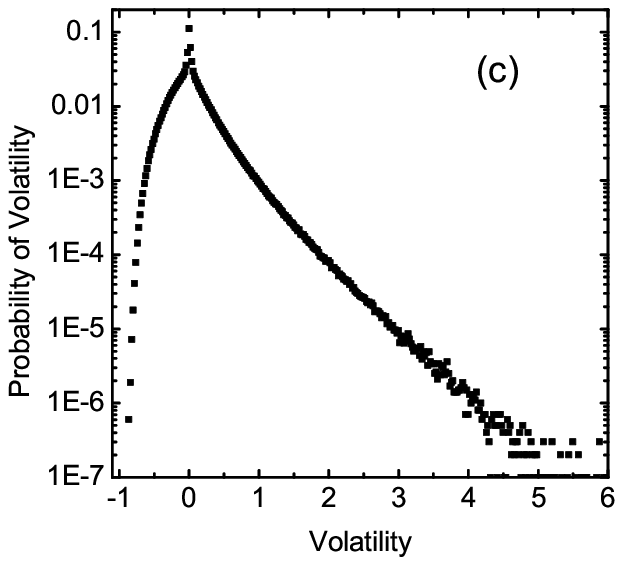}
\end{center}
\caption{The distribution of the volatility of agents' rank (a)
for the basic model, (b) for the model with uniform saving rate
$s=0.5$ and (c) for the  model with diverse saving rate.}
\label{fig:4}       
\end{figure}

To compare the mobilities quantitatively, we applied the
measurement index raised by G. S. Fields et al \cite{gfield}. The
mobility between the two sample recorded in different moments is
defined as
\begin{equation}\label{distance}
    l(t,t')= \displaystyle
    \frac{1}{N}\sum_{i=1}^{N}|\log(x_i(t))-\log(x_i(t'))|,
\end{equation}
where, $x_i(t)$ and $x_i(t')$ are the rank of agent $i$ at $t$ and
$t'$ respectively. It is obvious that the bigger the value of $l$,
the greater the degree of mobility. To eliminate the effect of the
randomness, we recorded more than 9000 samples continuously and
calculated the value of mobility $l$ between any two consecutive
samples. The average value of $l$s in these models are shown in
Table 1. It can be found that the degree of mobility decreases as
the saving rate increases in the model with uniform saving rate.
The intuition for this result is straightforward. The larger the
ratio agents put aside, the less money they take out to
participate the trade. Then, the less money to lose or win. Thus,
the higher saving rate, the less probability of change in rank or
mobility. The very low degree of mobility in the model with
diverse saving rate is due to its stratification.

To show more details of the mobility, we also obtain the
distribution of the volatility ($\frac{x_i(t')-x_i(t)}{x_i(t)}$)
which is shown in Fig.4. It is noted that the distributions of the
rank variety ratio are quite similar and follow power laws in the
basic model and the model with uniform saving rate. The exponent
of the power-law distribution is found to decrease as the saving
rate increases. This phenomenon is consistent with the alter trend
of the index because the higher the saving rate, the little money
is exchanged and the smaller the volatility of rank. Consequently,
when the saving rate increases, the right side of volatility
distribution will shift to the vertical axis, leading to a more
steeper tail. From Fig.4c, we can see that the volatility
distribution in the model with diverse saving rate ends with an
exponential tail as the times of simulations increase.

\section{Conclusion}

The dynamic phenomena of three transfer models, including money
circulation and economic mobility, are presented in this paper.
The holding time distributions in these models are demonstrated,
and the relation between the velocity of money and holding time of
money is expressed. Studies on this dynamic process lead us to a
good understanding the nature of money circulation process and
provide a new approach to the micro-foundation of the velocity.
The ``reranking" mobilities in these models are compared
graphically and quantitatively. This observation provide more
information about the dynamic mechanism behind the distribution.
Such investigations suggest that the characters of circulation and
mobility should be considered when constructing a multi-agent
model.

%
%

%

\begin{thebibliography}{99.}
%
%
%

\bibitem{model1} Dr\u{a}gulescu A, Yakovenko VM (2000) Statistical
mechanics of money. The European Physical Journal B 17:723--729

\bibitem{model2} Chakraborti A, Chakrabarti BK (2000) Statistical
mechanics of money: how saving propensity affects its
distribution. The European Physical Journal B 17:167--170

\bibitem{model3} Chatterjee A, Chakrabarti BK, Manna SS (2004)
Pareto law in a kinetic model of market with random saving
propensity. Physica A 335:155--163

\bibitem{solomon} Levy M, Solomon S (1996) Power laws are logarithmic Boltzmann laws.
International Journal of Modern Physics C 7:595--601

\bibitem{bouch} Bouchaud J-P, M\'{e}zard M (2000) Wealth condensation in a simple model of economy.
Physica A 282:536--545

\bibitem{sca} Scafetta N, Picozzi S, West BJ (2004) An out-of-equilibrium model of the distributions of
wealth. Quantitative Finance 4:353-364



\bibitem{pareto} Pareto V (1897) Cours d'Economie
Politique.  Macmillan, Paris

\bibitem{emp1} Levy M, Solomon S (1997) New evidence for the power-law distribution of wealth.
Physica A 242:90--94

\bibitem{emp2}  Dr\u{a}gulescu A, Yakovenko VM (2001) Evidence for the
exponential distribution of income in the USA. The European
Physical Journal B 20:585--589

\bibitem{emp3} Souma W (2001) Universal Structure of the personal income distribution.
Fractals 9:463--470

\bibitem{emp4} Souma W, Fujiwara Y, Aoyamac H, Kaizoji T, Aoki M (2003) Growth and fluctuations of personal income.
Physica A 321:598--604

\bibitem{emp5} Silva AC, Yakovenko VM (2005) Temporal evolution of the
``thermal" and ``superthermal" income classes in the USA during
1983--2001. Europhysics Letters 69:304--310



\bibitem{model4} Ding N, Wang Y, Xu J, Xi N (2004) Power-law distributions in
circulating money: effect of preferential behavior. International
Journal of Modern Physics B 18:2725--2729

\bibitem{follow} Hayes B (2002) Follow the money. American Scientist
90:400--405

\bibitem{slanina} Slanina F (2004) Inelastically scattering particles and wealth distribution in an open
economy. Physical Review E 69:046102

\bibitem{transfer} Wang Y, Ding N, Xi N (2005) Prospects of money transfer
models. In: Takayasu H (eds) Practical fruits of econophysics.
Springer, Tokyo


\bibitem{analysis1} Patriarca M, Chakraborti A, Kaski K (2004) Gibbs versus non-Gibbs distributions in money dynamics.
Physica A 340:334-339

\bibitem{analysis2} Patriarca M, Chakraborti A, Kaski K (2004) Statistical model with a standard $Gamma$ distribution.
Physical Review E 70:016104


\bibitem{analysis3} Repetowicz P, Hutzler S, Richmond P (2004) Dynamics of Money and Income Distributions.
arXiv: cond-mat/0407770



\bibitem{circulation1} Wang Y, Ding N, Zhang L (2003) The circulation of money
and holding time distribution. Physica A 324:665--677

\bibitem{circulation2} Ding N, Xi N, Wang Y (2003) Effects of saving and
spending patterns on holding time distribution. The European
Physical Journal B 36:149--153

\bibitem{velocity} Wang Y, Qiu H (2005) The velocity of money
in a life-cycle model. Physica A (in press)

\bibitem{mob} Ding N, Xi N, Wang Y (2005) The economic mobility in money transfer models. Submitted to Physica A.

\bibitem{kuznets} Kuznets SS (1966) Modern Economic Growth: Rate, Structure and Spread. Yale University, New Haven

\bibitem{jj} Jarvis S, Jenkins SP (1998) How much income mobility is there in Britain? The Economic Journal 108:1-16

\bibitem{gfield} Fields GS, Ok E (1999) Measuring Movement of Incomes. Economica 66:455-471




\end{thebibliography}
%



\printindex
\end{document}